\documentclass[12pt, a4paper]{article}

\usepackage[utf8]{inputenc}
\usepackage[english]{babel}

\usepackage{mathptmx} 
\usepackage{amsmath}

\usepackage{graphicx}
\usepackage[left=2.5cm, right=2.5cm, top=2.5cm, bottom=2.5cm]{geometry}

\usepackage{setspace}
\usepackage{ragged2e} 
\usepackage{booktabs}
\usepackage{array}
\usepackage{adjustbox}
\usepackage{longtable}
\usepackage{tabularx}
\usepackage{tikz}
\usetikzlibrary{positioning,arrows.meta}

\usepackage{enumitem}

\usepackage{xcolor}

\usepackage{hyperref}
\hypersetup{
    colorlinks=true,
    linkcolor=blue,
    urlcolor=cyan,
    pdftitle={The Effects of IMF Programs},
    pdfauthor={Synthetic Analysis}
}

\usepackage{natbib}
\newcolumntype{L}[1]{>{\RaggedRight\arraybackslash}p{#1}}
\newcolumntype{Y}{>{\centering\arraybackslash}X}

\title{\textbf{The effects of International Monetary Fund programs: a systematic review with narrative synthesis on poverty, inequality, and social indicators}}
\author{Ricardo Alonzo Fernández Salguero}
\date{\today}

\begin{document}

\maketitle

\begin{abstract}
\noindent \textbf{Background:} The debate over the impact of the International Monetary Fund's (IMF) structural adjustment programs on borrowing countries has been a central topic in political economy for decades. The empirical evidence is profoundly heterogeneous and often contradictory.
\textbf{Objective:} This systematic review with narrative synthesis synthesizes and critically analyzes the literature to assess the effects of such programs on key indicators like poverty, income inequality, social spending (health and education), health outcomes (mortality, infectious diseases), the labor market, and informality.
\textbf{Methods:} A systematic search was conducted in five academic databases and grey literature, following PRISMA guidelines. A total of 53 studies and empirical contributions that met predefined selection criteria were included. The risk of bias for each study was assessed, paying special attention to the control of endogeneity and selection bias. Due to the methodological heterogeneity of the studies, the results were synthesized narratively.
\textbf{Results:} A minority of studies, often employing methodologies with a higher risk of bias such as Propensity Score Matching (PSM), find no significant negative effects. In contrast, a robust body of research, using quasi-experimental designs with a low risk of bias like Instrumental Variables (IV), documents an increase in inequality, a deterioration of health outcomes (especially in tuberculosis and child mortality), and a rise in the informal economy as a consequence of IMF conditionalities.
\textbf{Conclusion:} The differences in findings are largely attributed to the diversity of methodologies and the rigor with which they address causal inference. Higher-quality evidence suggests that IMF programs, particularly those with austerity conditionalities and structural reforms, impose significant social costs. These impacts call for a redesign of conditionality that prioritizes social protection to meet the Sustainable Development Goals.
\end{abstract}

\section{Introduction}

The International Monetary Fund (IMF), since its creation at Bretton Woods, has played a central role in the global financial architecture, primarily through the provision of loans to countries facing balance of payments crises. These loans, however, are not unconditional; they are accompanied by a set of economic policy reforms, known as conditionalities, which aim to stabilize the macroeconomy and lay the groundwork for sustainable growth \citep{joyce2004}. The nature and rigor of these conditionalities have evolved, from the emphasis on austerity and liberalization of the "Washington Consensus" to a more recent discourse that incorporates the importance of social spending and inclusive growth. Nevertheless, in practice, structural adjustment policies have historically included measures of fiscal consolidation, trade and financial sector liberalization, privatization of state-owned enterprises, and labor market reforms \citep{andritzky2021}.

The effectiveness and consequences of these programs have been the subject of intense academic and political debate. Supporters of the IMF argue that its interventions are a necessary, albeit sometimes bitter, remedy to correct unsustainable macroeconomic imbalances and that, in the long run, they promote stability and growth \citep{doroodian1994, balima2020}. Although \cite{balima2020} reports publication bias, they do not correct for it in their meta-analysis, which could change their conclusions. From this perspective, any negative social impact is considered a transitory and unavoidable cost of a necessary adjustment, or it is attributed to the pre-existing crisis that motivated the intervention and not to the program itself \citep{eicher2024}. On the other hand, a vast critical literature argues that IMF policies, especially austerity measures, impose disproportionate social costs, exacerbating poverty and inequality, weakening health and education systems, and harming the most vulnerable populations \citep{stuckler2009, lang2016, stubbs2021}.

This review delves into this complex and polarized debate with a distinctive approach. Unlike previous reviews, which often focus on a single sector such as health \citep{thomson2017} or on purely macroeconomic outcomes, this research offers several key contributions. First, it provides a multidimensional scope by synthesizing evidence across a broad spectrum of interconnected social outcomes: poverty, inequality, social spending, health, mortality, labor market, and informality. Second, it presents an updated and comprehensive coverage of the literature, incorporating the most recent studies and allowing for a view of the evolution of conditionality and its evaluation over time. Finally, and crucially, it uses the comparative analysis of research methodologies as the primary lens to explain the divergences in findings, arguing that rigor in addressing causal inference is a key predictor of the studies' conclusions.

By grouping and comparing studies not only by their conclusions but also by their methodological designs and quality assessment, this review aims to offer a nuanced overview that recognizes the contingency of the IMF's effects and the inherent complexities in their evaluation.

\section{Review Methodology}

This research is structured as a systematic review with a narrative synthesis, an approach designed to map, interpret, and critically synthesize the diverse and often methodologically heterogeneous evidence on the socioeconomic impacts of IMF programs. To ensure transparency and rigor, the reporting process adhered to the PRISMA 2020 (Preferred Reporting Items for Systematic Reviews and Meta-Analyses) checklist guidelines \citep{page2021}. Given that the intervention measures (types of programs and conditionalities) and outcomes (different social indicators) vary considerably among studies, a quantitative meta-analysis was not feasible. Instead, a narrative synthesis was chosen, which allows for contextualizing contradictory findings, with special attention to the methodological designs underlying each study's conclusions.

\subsection{Search and Identification Strategy}

A systematic literature search was conducted during the period from July 5, 2025, to October 16, 2025. The search covered five top-tier academic databases to ensure multidisciplinary coverage: Scopus and Web of Science for their broad scope; PubMed for its specialization in public health; EconLit for its focus on economic literature; and Google Scholar to capture additional studies and grey literature. A sensitive Boolean search strategy was designed and applied, adapted to the syntax of each database, combining key terms related to the intervention and outcomes of interest. The search logic was as follows:

\begin{quote}
\small
\begin{verbatim}
("International Monetary Fund" OR "IMF") 
AND ("structural adjustment" OR "conditionality" 
OR "austerity" OR "IMF program*") 
AND ("poverty" OR "inequality" OR "social spending" 
OR "health outcome*" OR "education" OR "infant mortality" 
OR "maternal mortality" OR "labor market" 
OR "informal economy" OR "gender gap")
\end{verbatim}
\end{quote}

To complement the database search, a manual search of grey literature was performed, including working papers from the IMF, the World Bank, and reports from relevant non-governmental organizations (e.g., Oxfam, Center for Economic and Policy Research). Additionally, a "snowballing" technique was applied, systematically reviewing the reference lists of key articles and systematic reviews identified to capture further studies. This exhaustive search process yielded an initial total of 845 records.

\subsection{Selection Criteria and Screening}

The study selection process was guided by a set of predefined and rigorous inclusion and exclusion criteria, which are summarized in Table \ref{tab:criterios_seleccion}. The goal was to include only empirical studies that directly evaluated the impact of IMF programs on social indicators in low- and middle-income countries. Purely theoretical studies, commentaries, and analyses focused exclusively on macroeconomic outcomes without an explicit link to social welfare were excluded.

\begin{figure}[htbp]
\centering
\begin{tikzpicture}[
    node distance = 1.7cm and 3cm, 
    every node/.style = {font=\small},
    box/.style = {
        rectangle,
        draw,
        fill=blue!8,
        text width=6cm,
        align=center,
        minimum height=1.4cm
    },
    sidebox/.style = {
        rectangle,
        draw,
        text width=6cm,
        align=left,
        minimum height=1.4cm
    },
    arrow/.style = {-latex, thick}
]

\node[box] (identification)
  {\textbf{Identification: database search}\\
   Records identified (n = 845)};

\node[box, below=of identification] (screening)
  {\textbf{Screening}\\
   Records after duplicates removed (n = 724)};

\node[box, below=of screening] (screening2)
  {Records screened by title and abstract (n = 724)};

\node[box, below=of screening2] (eligibility)
  {\textbf{Eligibility}\\
   Full-text reports assessed for eligibility (n = 220)};

\node[box, below=2.5cm of eligibility] (included)
  {\textbf{Inclusion}\\
   Studies included in narrative synthesis (n = 53)};

\node[sidebox, right=of screening] (excluded1)
  {Duplicate records removed (n = 121)};

\node[sidebox, right=of screening2] (excluded2)
  {Records excluded for thematic irrelevance (n = 504)};

\node[sidebox, right=of eligibility] (excluded3)
  {\textbf{Full-text reports excluded (n = 167), with reasons:}\\[-0.4em]
   \begin{itemize}[leftmargin=1.2em, topsep=1pt, itemsep=2pt, parsep=0pt]
     \item Outcome of interest not analyzed (n = 95)
     \item Non-empirical methodology (e.g., editorial) (n = 42)
     \item Full text unavailable (n = 30)
   \end{itemize}
  };

\draw[arrow] (identification.south) -- (screening.north);
\draw[arrow] (screening.south) -- (screening2.north);
\draw[arrow] (screening2.south) -- (eligibility.north);
\draw[arrow] (eligibility.south) -- (included.north);

\draw[arrow] (screening.east) -- ++(0.25,0) |- (excluded1.west);
\draw[arrow] (screening2.east) -- ++(0.25,0) |- (excluded2.west);
\draw[arrow] (eligibility.east) -- ++(0.25,0) |- (excluded3.west);

\end{tikzpicture}
\caption{PRISMA flow diagram of the study selection process.}
\label{fig:prisma_flow}
\end{figure}
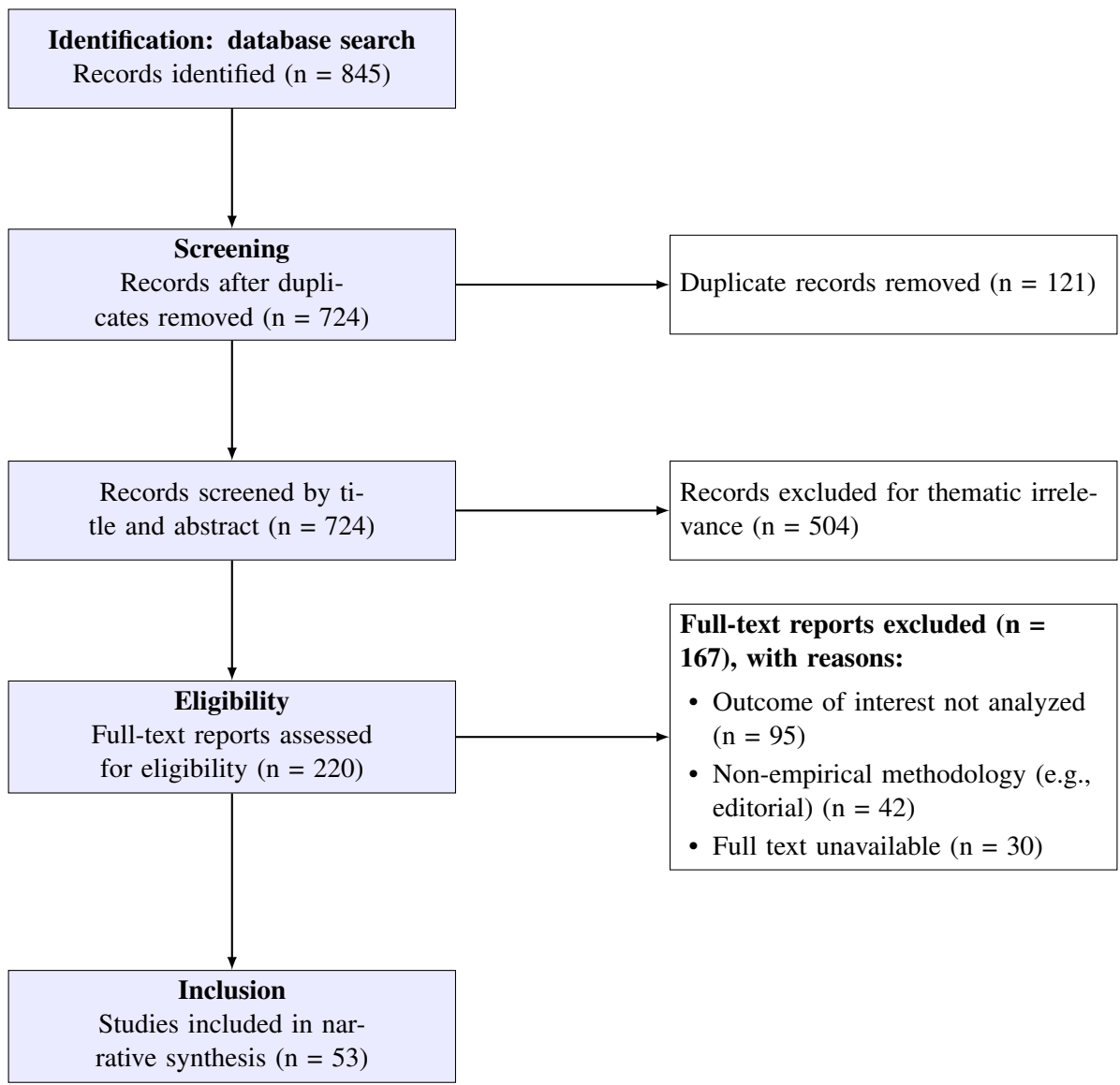

The screening of records was conducted in two phases by two independent reviewers, with any disagreements resolved through discussion and consensus. In the first phase, the titles and abstracts of 724 unique records were evaluated after removing duplicates. In the second phase, the full text of 220 articles that appeared potentially eligible was retrieved and assessed to confirm they met all criteria. This rigorous process resulted in the final inclusion of 53 studies and empirical contributions for the narrative synthesis. The complete flow of the process is detailed in the PRISMA diagram (Figure \ref{fig:prisma_flow}).

\begin{table}[htbp]
  \centering
  \caption{Inclusion and exclusion criteria for studies}
  \label{tab:criterios_seleccion}
  \begin{adjustbox}{width=\textwidth}
  \begin{tabular}{L{7.5cm} L{7.5cm}}
    \toprule
    \textbf{Inclusion Criteria} & \textbf{Exclusion Criteria} \\
    \midrule
    \textbf{Population/Context:} Low- and middle-income countries that have participated in at least one IMF program. & \textbf{Population/Context:} Studies focused exclusively on high-income countries. \\
    \addlinespace
    \textbf{Intervention:} Participation in a structural adjustment program or conditional loan from the IMF. & \textbf{Intervention:} Analysis of other international financial institutions without direct comparison or mention of the IMF. \\
    \addlinespace
    \textbf{Outcomes:} The study must empirically analyze at least one of the following: poverty, inequality (income, gender), social spending (health, education), health outcomes (mortality, morbidity), labor market (informality, rights). & \textbf{Outcomes:} Studies focused solely on macroeconomic outcomes without an explicit link to social indicators. \\
    \addlinespace
    \textbf{Study Type:} Quantitative empirical studies (panel, time series, quasi-experimental), qualitative case studies, systematic reviews, and meta-analyses. & \textbf{Study Type:} Editorials, letters, commentaries, opinion pieces without primary or secondary data analysis. \\
    \bottomrule
  \end{tabular}
  \end{adjustbox}
\end{table}

\subsection{Data Extraction, Quality Assessment, and Synthesis}

Data extraction from the 53 included studies was performed using a standardized template. To assess the quality and risk of bias of quantitative studies, criteria from tools like ROBINS-I for non-randomized studies were adapted. Each study was evaluated in three key domains for the validity of its conclusions. First, control for selection bias and endogeneity was analyzed, assessing whether the study employed robust methods such as instrumental variables (IV), difference-in-differences (DiD), or regression discontinuity designs (RDD) to address the fundamental problem that countries participating in IMF programs are systematically different from those that do not. Second, measurement quality was evaluated, examining whether program participation and social outcome variables were reliably defined and measured. Third, control for confounding factors was assessed, verifying whether the statistical model included appropriate control variables for other plausible influences on social outcomes.

Based on this evaluation, studies were classified with a \textit{low}, \textit{moderate}, or \textit{high} risk of bias. A detailed list of the 53 included studies and their quality assessment is presented in the Appendix (Table \ref{tab:apendice_estudios}). The evidence was synthesized narratively, grouping studies by theme and contrasting findings in light of their methodological quality, which allowed for identifying patterns where the rigor of the research design correlates with the direction and magnitude of the reported effects.

\section{Results}

\subsection{Effects on Poverty and Inequality}

The impact of IMF programs on poverty and income inequality is one of the most contentious areas in the literature, with findings that vary drastically. This divergence, however, is not random; it aligns strongly with the methodological rigor employed to address problems of causal inference.

One stream of research, often characterized by a moderate to high risk of methodological bias, suggests that IMF programs are not systematically detrimental to income distribution. A representative example is the study by \citet{bird2020}, which uses propensity score matching (PSM) and finds no significant increase in poverty or inequality. However, PSM can only control for observable differences between countries that participate in programs and those that do not, leaving the potential bias from unobservable factors, such as political will for reform or the underlying severity of the crisis, unresolved. This line of findings is consistent with the IMF's institutional position, which argues that its more recent programs, such as those supported by the Poverty Reduction and Growth Facility (PRGF), have evolved to incorporate social safeguards and a focus on pro-poor spending \citep{gupta2003, shirai2004}. Nonetheless, the effectiveness of these safeguards has been questioned by external analyses, which describe them as insufficient or, at worst, as a "fig leaf for austerity" \citep{kentikelenis2023_floors, stubbs2017}.

In stark contrast, an overwhelming number of studies, especially those with a low risk of bias that employ quasi-experimental designs, find that IMF programs exacerbate income inequality. \citet{lang2016}, in a methodologically robust study that uses IMF liquidity as an instrumental variable to isolate exogenous variation in program participation, concludes that these programs increase inequality primarily by causing absolute income losses for the poorest strata. Similarly, \citet{stubbs2021}, using an IV design and an original dataset on IMF fiscal austerity targets, demonstrate that stricter fiscal consolidation is associated with greater inequality, by concentrating income in the top 10\%, and with higher poverty rates. Table \ref{tab:poverty_inequality} summarizes these contrasting findings, showing how more rigorous methods tend to find negative effects.

The causal mechanism through which these regressive distributional effects operate is increasingly well-documented. \citet{biglaiser2022} show that structural reforms, rather than stabilization measures, increase poverty by raising unemployment and reducing government revenues available for transfers. \citet{easterly2001, easterly2003} offer a nuanced perspective, arguing that adjustment programs reduce the "poverty-reducing elasticity of growth," meaning that the poor benefit less from the economic growth that occurs under IMF tutelage.

The impact of these programs is also profoundly asymmetric in terms of gender. High-quality evidence shows that IMF programs not only deteriorate women's economic rights in general \citep{detraz2015}, but also increase gender gaps in the labor market, such as unemployment and labor force participation \citep{kern2024}. Austerity policies and fiscal reforms, such as increases in value-added taxes (VAT), disproportionately fall on women due to their predominant roles in unpaid care work and their consumption patterns, which are more focused on essential goods \citep{donno2024, abdo2019}.

\begin{table}[htbp]
  \centering
  \caption{Synthesis of key studies on the impact of the IMF on poverty and inequality}
  \label{tab:poverty_inequality}
  \begin{adjustbox}{width=\textwidth}
  \begin{tabular}{L{3.5cm} L{6.5cm} L{5cm}}
    \toprule
    \textbf{Authors} & \textbf{Main Finding} & \textbf{Methodology and Risk of Bias} \\
    \midrule
    \citet{bird2020} & Programs are not associated with an increase in poverty or inequality. & PSM (Moderate risk of bias, does not control for unobservables). \\
    \addlinespace
    \citet{lang2016} & Programs increase income inequality, mainly due to income losses for the poor. & Instrumental Variable (IV) based on IMF liquidity (Low risk of bias). \\
    \addlinespace
    \citet{stubbs2021} & Stricter austerity increases inequality and poverty. & Longitudinal analysis with IV and continuous DiD (Low risk of bias). \\
    \addlinespace
    \citet{easterly2001, easterly2003} & They reduce the poverty-reducing elasticity of growth. & Regression with IV (Moderate to low risk of bias). \\
    \addlinespace
    \citet{biglaiser2022} & Structural reforms increase poverty. & Instrumental model (IV-2SLS) (Low risk of bias). \\
    \addlinespace
    \citet{kern2024} & They increase the gender unemployment gap. & Error correction model with IV (Low risk of bias). \\
    \bottomrule
  \end{tabular}
  \end{adjustbox}
\end{table}

\subsection{Impact on Social Spending, Health, and Education}

The logic of fiscal austerity, central to many IMF programs, suggests that to reduce deficits, governments will cut public spending, often affecting social sectors. The empirical evidence on this point is, again, divided, and its interpretation depends crucially on the methodological quality of the studies.

On one hand, a series of studies, often affiliated with the IMF or using methodologies with a moderate risk of bias, argue that programs protect or even increase social spending. \citet{clements2013}, using a system GMM model, conclude that programs have a positive and significant effect on health and education spending in low-income countries. Similarly, \citet{gupta2020} and a recent report on Sub-Saharan Africa \citep{afavi2024} suggest that conditionality in public finance management can, in the long term, improve social spending and associated outcomes.

However, a vast counter-literature, often based on more robust research designs for causal inference, documents pronounced negative effects. \citet{kentikelenis2023}, with a sophisticated composite instrumental variable design (low risk of bias), find that each additional structural condition imposed by the IMF reduces government spending on education. Other qualitative and quantitative studies document how austerity policies have led to decades of underinvestment and how wage bill ceilings imposed by the IMF directly limit the ability of governments to hire essential health and education professionals \citep{marphatia2010, rowden2010, kolko2020}.

These cuts and restrictions have direct and often devastating consequences for public health. Table \ref{tab:health_outcomes} summarizes some of the most consistent and alarming findings. The association between IMF programs and the deterioration of tuberculosis indicators is one of the most robust and replicated results in the literature, with studies showing significant increases in the incidence, prevalence, and mortality of the disease in countries under adjustment programs \citep{stuckler2008, murray2008, maynard2012}. Beyond TB, programs have been linked to a general weakening of health systems \citep{stuckler2009, sobhani2019} and an increase in mortality from infectious diseases in general \citep{nosrati2021_infectious}.

Maternal and child health is another area where the negative impact has been convincingly documented. \citet{thomson2017}, in a systematic review, conclude that structural adjustment programs have a detrimental impact in this area. Specific studies, using panel models with fixed effects, confirm this association with higher rates of infant \citep{shandra2012} and maternal \citep{pandolfelli2014} mortality in Sub-Saharan Africa. Even more forcefully, \citet{nosrati2021_child}, employing a robust instrumental variables methodology (low risk of bias), estimates that IMF programs cause a statistically significant excess of deaths of children under five. The criticism extends to the IMF's own narratives, whose claims about their positive contributions to public health have been systematically refuted by independent evaluations that point out contradictions with the available evidence \citep{stuckler2010}.

\begin{table}[htbp]
  \centering
  \caption{Synthesis of studies on the impact of the IMF on health and social spending}
  \label{tab:health_outcomes}
  \begin{adjustbox}{width=\textwidth}
  \begin{tabular}{L{3.5cm} L{7cm} L{4.5cm}}
    \toprule
    \textbf{Authors} & \textbf{Main Finding} & \textbf{Methodology and Risk of Bias} \\
    \midrule
    \multicolumn{3}{l}{\textit{\textbf{Positive/neutral findings on social spending}}} \\
    \citet{clements2013} & Programs increase education and health spending in low-income countries. & System GMM model (Moderate risk of bias). \\
    \addlinespace
    \citet{boachie2022} & In Ghana, programs did not significantly reduce health spending. & Three-stage least squares (3SLS) (Moderate risk of bias). \\
    \midrule
    \multicolumn{3}{l}{\textit{\textbf{Negative findings on social spending and health}}} \\
    \citet{kentikelenis2023} & Each additional structural condition reduces education spending. & Composite IV (Low risk of bias). \\
    \addlinespace
    \citet{stuckler2008} & 16.6\% increase in tuberculosis mortality in post-communist countries. & Panel regression with fixed effects (Moderate risk of bias). \\
    \addlinespace
    \citet{thomson2017} & Detrimental impact on maternal and child health. & Systematic-narrative review (Quality dependent on primary studies). \\
    \addlinespace
    \citet{nosrati2021_child} & Programs cause an excess of up to 90 deaths of children under five per 1,000 live births. & Panel regressions with IV (Low risk of bias). \\
    \bottomrule
  \end{tabular}
  \end{adjustbox}
\end{table}

\subsection{Consequences for the Labor Market and Informality}

IMF conditionalities often include labor market reforms aimed at increasing flexibility, such as facilitating dismissals and reducing the public sector. While the stated goal is to stimulate formal job creation, a growing body of evidence suggests that these reforms can have the unintended consequence of expanding the informal economy. \citet{ohnsorge2022} point out that widespread informality is associated with worse economic outcomes and higher poverty. Recent studies with robust methodologies (low risk of bias) confirm that IMF policies contribute to its expansion. \citet{adam2024}, using an augmented inverse probability weighting (IPWRA) technique, find that reductions in public sector employment, a common conditionality, lead to an increase in the informal economy of 1.3 percentage points of GDP five years later. \citet{chletsos2020_informal} corroborate this finding, showing that both program participation and its structural conditionalities increase the size of the shadow economy.

This growth of informality has serious implications for worker welfare and public health. Informal workers, by definition, lack formal contracts, social protection, and access to social security. This translates into poorer access to health services \citep{naicker2021} and, as an ecological study in Latin America found, higher mortality rates \citep{silva-penaherrera2021}. Informality creates a vicious cycle of vulnerability that adjustment policies seem to exacerbate.

In addition to fostering informality, the reforms have been criticized for directly weakening labor rights. \citet{lee2021} conclude that IMF programs have overall negative effects on workers' rights, both de jure and de facto, although they note that domestic politics can mediate these effects. In contexts of economic crisis and adjustment, these pressures can also lead to an increase in the worst forms of labor, including child labor, as a household survival strategy in the face of reduced income and the erosion of social safety nets \citep{mark2021}. Table \ref{tab:labor_informality} summarizes these key findings.

\begin{table}[htbp]
  \centering
  \caption{Synthesis of studies on the impact of the IMF on the labor market and informality}
  \label{tab:labor_informality}
  \begin{adjustbox}{width=\textwidth}
  \begin{tabular}{L{3.5cm} L{7.5cm} L{4cm}}
    \toprule
    \textbf{Authors} & \textbf{Main Finding} & \textbf{Methodology and Risk of Bias} \\
    \midrule
    \citet{adam2024} & Public employment reductions increase the informal economy by 1.3 p.p. of GDP. & Dynamic IPWRA (Low risk of bias). \\
    \addlinespace
    \citet{chletsos2020_informal} & Program participation and its conditionalities increase the shadow economy. & Panel analysis with endogeneity controls (Low risk of bias). \\
    \addlinespace
    \citet{lee2021} & Negative effects on labor rights, exacerbated by strict conditions. & Qualitative-comparative analysis (Not applicable). \\
    \addlinespace
    \citet{mark2021} & Greater compliance with reforms is associated with an increase in child labor. & Selection model with control function (Low risk of bias). \\
    \bottomrule
  \end{tabular}
  \end{adjustbox}
\end{table}

\subsection{Comparative Analysis of Methodologies}

The marked divergence in the literature's findings cannot be understood without a critical analysis of the methodologies employed. Assessing the causal impact of IMF programs is a formidable econometric challenge due to two main problems: selection bias (countries that turn to the IMF are not a random sample, but are already in a deep crisis) and endogeneity (country characteristics can influence both the likelihood of receiving a program and social outcomes). The way researchers address these problems decisively influences their conclusions.

Early studies, often based on panel regressions with fixed effects, could identify correlations but struggled to establish causality, being classified with a moderate risk of bias \citep{stuckler2008}. To overcome this, more sophisticated methods were adopted. Propensity Score Matching (PSM), used by \citet{bird2020}, attempts to create a synthetic control group, but its main limitation is that it can only control for observable differences, leaving a potential residual bias due to unobservable factors. Similarly, Dynamic Panel Models (GMM), such as those employed by \citet{clements2013}, address certain types of endogeneity but can be sensitive to model specification and the choice of instruments.

The methodologies considered most robust for causal inference in this field are those that use quasi-experimental designs, with the Instrumental Variables (IV) approach being prominent. This method seeks a variable (the instrument) that is correlated with participation in an IMF program but does not directly affect the outcome of interest (e.g., inequality) other than through its effect on participation. Studies that successfully implement this approach, such as those by \citet{lang2016}, \citet{stubbs2021}, and \citet{nosrati2021_child}, are classified with a low risk of bias and consistently find the strongest and most statistically significant negative effects. This suggests that less rigorous methods may be underestimating the adverse impacts by failing to fully separate the effect of the program from the effect of the pre-existing crisis. In summary, there is a clear hierarchy of evidence where studies with the strongest causal identification designs tend to converge on the conclusion that IMF programs have significant social costs. Table \ref{tab:method_comparison} summarizes this comparison.

\begin{longtable}{L{3cm} p{5cm} p{6cm}}
\caption{Comparison of methodologies and their implications for results} \label{tab:method_comparison} \\
\toprule
\textbf{Methodology} & \textbf{Description and advantages} & \textbf{Limitations and typical findings} \\
\midrule
\endfirsthead
\multicolumn{3}{c}%
{{\bfseries \tablename\ \thetable{} -- continued from previous page}} \\
\toprule
\textbf{Methodology} & \textbf{Description and Advantages} & \textbf{Limitations and Typical Findings} \\
\midrule
\endhead
\midrule \multicolumn{3}{r}{{Continued on next page}} \\
\endfoot
\bottomrule
\endlastfoot
\textbf{Panel Regression (Fixed Effects)} & Controls for time-invariant characteristics. Useful for identifying correlations. \citep{stuckler2008} & \textbf{Limitations:} Vulnerable to selection bias and endogeneity. Causality is difficult to establish. \textbf{Findings:} Mixed results, often negative associations. (Moderate risk of bias). \\
\addlinespace
\textbf{Propensity Score Matching (PSM)} & Creates a statistical control group based on observable characteristics. \citep{bird2020} & \textbf{Limitations:} Does not control for unobservable factors. \textbf{Findings:} Tends to find smaller or null effects. (Moderate to high risk of bias). \\
\addlinespace
\textbf{Generalized Method of Moments (GMM)} & Used in dynamic panels to address the endogeneity of lagged variables. \citep{clements2013} & \textbf{Limitations:} Sensitive to the choice of instruments. \textbf{Findings:} Often used in IMF studies that find positive or neutral effects. (Moderate risk of bias). \\
\addlinespace
\textbf{Instrumental Variables (IV) / 2-Stage OLS} & Isolates exogenous variation in program participation for stronger causal inference. \citep{lang2016, stubbs2021, nosrati2021_child} & \textbf{Limitations:} The challenge is finding a valid instrument. \textbf{Findings:} Generally finds the largest and most significant negative effects. (Low risk of bias). \\
\addlinespace
\textbf{Systematic Review} & Synthesizes the results of multiple primary studies. \citep{thomson2017} & \textbf{Limitations:} The conclusion depends on the quality of the included studies. \textbf{Findings:} Reflects the consensus of evidence in an area. (Variable quality). \\
\end{longtable}

\section{Discussion and Conclusion}

This systematic review of a vast and diverse literature on the effects of IMF programs reveals a complex and fractured landscape. Far from offering a simple verdict, the accumulated evidence underscores that there is no monolithic "IMF effect." However, despite the heterogeneity, consistent patterns emerge when the evidence is analyzed through the prism of methodological rigor. The highest-quality empirical evidence, from studies with robust quasi-experimental designs, consistently points to adverse social consequences. On average, IMF programs are associated with an increase in income inequality, a deterioration of public health, and an expansion of labor informality. The claim that programs are benign or beneficial is supported by a minority of studies, often with greater methodological limitations for establishing causality.

The mechanisms through which these negative effects operate are increasingly well-identified. Fiscal austerity, a central pillar of many programs, leads to cuts in social spending that erode the provision of essential public services. Labor market reforms, aimed at flexibility, often translate into greater precarity and less protection for workers. These effects are not gender-neutral; women often bear a disproportionate burden of the adjustment costs. Methodological choice is of paramount importance: the discrepancy between the findings of different studies is largely the result of applying different tools to unravel a complex causal problem.

\subsection{Policy Implications and Future Research Directions}

The findings of this review have direct and urgent implications for policy reform. It is imperative that the design of IMF programs evolves to internalize their social costs. This requires the mandatory inclusion of ex-ante social and gender impact assessments, conducted independently, to anticipate and mitigate adverse effects on vulnerable populations. Reforming fiscal conditionality is another crucial point; "social spending floors" must be binding, transparent, and designed in consultation with civil society, and they must be accompanied by the elimination of indiscriminate wage bill ceilings in the health and education sectors. Furthermore, labor conditionality should be reoriented to promote formalization and decent work, rather than a flexibility that leads to precarity. Finally, to enable independent and rigorous evaluation, the IMF must radically increase its transparency, making all program documents and underlying data publicly available.

This review also reveals research gaps that need to be addressed. More studies are needed that use robust causal designs to analyze long-term effects, beyond the first few post-program years, as structural impacts may take longer to materialize. It is essential to further investigate the heterogeneity of effects to understand why outcomes vary so much between countries, focusing on the mediating role of national political institutions, such as the quality of democracy and state capacity. Likewise, it is crucial to assess the social impact of the IMF's new lending facilities, such as the Resilience and Sustainability Trust (RST), to determine whether they replicate or overcome the problems of traditional conditionality.

The accumulated evidence suggests that, for many countries, the IMF's "medicine" has had serious side effects that often worsen the patient's social and health condition. Ignoring these costs is not only an analytical omission but also a profound issue of global distributive justice that requires urgent attention and fundamental reform.



\clearpage
\appendix
\section{Studies Included in the Systematic Review}

\begin{longtable}{p{4.5cm} p{7.5cm} l}
\caption{List and quality assessment of the 53 studies and empirical contributions included} \label{tab:apendice_estudios} \\
\toprule
\textbf{Reference} & \textbf{Main focus and key finding} & \textbf{Risk of bias} \\
\midrule
\endfirsthead
\multicolumn{3}{c}%
{{\bfseries \tablename\ \thetable{} -- continued from previous page}} \\
\toprule
\textbf{Reference} & \textbf{Main focus and key finding} & \textbf{Risk of bias} \\
\midrule
\endhead
\midrule \multicolumn{3}{r}{{Continued on next page}} \\
\endfoot
\bottomrule
\endlastfoot
\citet{abdo2019} & Gender and inequality in MENA. IMF policies worsen gender inequality and poverty. & Qualitative \\
\addlinespace
\citet{adam2024} & Informality. IMF public employment reductions increase the informal economy. & Low \\
\addlinespace
\citet{afavi2024} & Social spending in Africa. IMF programs preserve or increase social spending. & Moderate \\
\addlinespace
\citet{apodaca2017} & Child mortality. IMF loans have mixed and sometimes negative effects. & Moderate \\
\addlinespace
\citet{baker2010} & AIDS pandemic. IMF policies weakened health systems, exacerbating the crisis. & Qualitative \\
\addlinespace
\citet{balima2020} & Economic growth. Meta-analysis shows highly variable effects, context-dependent. & Review \\
\addlinespace
\citet{bas2013} & Economic growth. Programs have positive effects once adverse selection is controlled for. & Moderate \\
\addlinespace
\citet{beste2016} & Poverty and health in Mozambique. IMF policies have had a negative impact. & Qualitative \\
\addlinespace
\citet{biglaiser2022} & Poverty. IMF structural conditions increase poverty. & Low \\
\addlinespace
\citet{bird2020} & Poverty and inequality. No systematic negative effects are found. & Moderate \\
\addlinespace
\citet{boachie2022} & Health spending in Ghana. IMF programs have no significant effect on reducing spending. & Moderate \\
\addlinespace
\citet{chletsos2020_inequality} & Income inequality. IMF programs increase inequality. & Moderate \\
\addlinespace
\citet{chletsos2020_informal} & Informality. IMF programs and conditionalities increase the informal economy. & Low \\
\addlinespace
\citet{clements2013} & Social spending. Programs increase health and education spending in low-income countries. & Moderate \\
\addlinespace
\citet{coburn2015} & Child mortality in Africa (AfDB). Adjustment loans increase mortality, investment loans reduce it. & Moderate \\
\addlinespace
\citet{daoud2019} & Child poverty. IMF programs increase the probability of child poverty by 14\%. & Low \\
\addlinespace
\citet{demir2022} & Economic complexity. IMF programs have no effect on export structure or diversification. & Low \\
\addlinespace
\citet{detraz2015} & Women's rights. IMF programs deteriorate women's economic rights. & Moderate \\
\addlinespace
\citet{donno2024} & Gender and taxes. IMF tax advice (VAT) has unintended negative effects on women. & Low \\
\addlinespace
\citet{doroodian1994} & Growth and inflation. Programs improve macroeconomic indicators in the short term. & High \\
\addlinespace
\citet{easterly2001} & Poverty. Programs reduce the effectiveness of growth in reducing poverty. & Moderate \\
\addlinespace
\citet{eicher2024} & Gender inequality. No systematic differences found in gender indicators. & Moderate \\
\addlinespace
\citet{ferre2016} & Health in Latin America. Countries without IMF loans have better health indicators. & Moderate \\
\addlinespace
\citet{gupta2020} & Social spending. Structural conditionality can boost health and education spending in the long run. & Moderate \\
\addlinespace
\citet{isran2014} & Social cost in Pakistan. IMF austerity had negative impacts on health, education, and employment. & Mixed \\
\addlinespace
\citet{joyce2004} & Economic impact. Review shows programs improve the external balance but do not attract capital. & Review \\
\addlinespace
\citet{kentikelenis2023} & Education spending. Each additional IMF condition reduces education spending. & Low \\
\addlinespace
\citet{kern2024} & Gender and labor market. IMF programs increase the gender unemployment gap. & Low \\
\addlinespace
\citet{lang2016} & Income inequality. Programs increase inequality by harming the poor. & Low \\
\addlinespace
\citet{lee2021} & Labor rights. Programs have negative effects on labor rights. & Qualitative \\
\addlinespace
\citet{mark2021} & Child labor. Greater compliance with IMF reforms is associated with an increase in child labor. & Low \\
\addlinespace
\citet{marphatia2010} & Health and education workforce. IMF policies limit hiring and investment. & Qualitative \\
\addlinespace
\citet{martinez-chiluisa2022} & Inequality. Non-concessional IMF loans increase inequality. & Moderate \\
\addlinespace
\citet{maynard2012} & Tuberculosis. Structural adjustment loans are detrimental to TB control. & Moderate \\
\addlinespace
\citet{murray2008} & Health (Tuberculosis). IMF loans are associated with worse TB outcomes in Eastern Europe. & Moderate \\
\addlinespace
\citet{nosrati2021_child} & Child mortality. IMF programs cause a significant excess of child deaths. & Low \\
\addlinespace
\citet{nosrati2021_infectious} & Infectious disease mortality. Adjustment programs increase mortality. & Low \\
\addlinespace
\citet{ozturk2011} & Macroeconomics in Latin America. Positive effects on balance of payments but negative on GDP and consumption. & Moderate \\
\addlinespace
\citet{pandolfelli2014} & Maternal mortality in Africa. Adjustment loans are associated with higher maternal mortality. & Moderate \\
\addlinespace
\citet{reinsberg2022} & Distributive consequences. Governments protect their supporters and impose costs on the opposition. & Low \\
\addlinespace
\citet{rowden2010} & Inflation and social spending. The IMF prioritizes low inflation over social spending and growth. & Qualitative \\
\addlinespace
\citet{shandra2012} & Child mortality in Africa. IMF structural adjustment is associated with higher child mortality. & Moderate \\
\addlinespace
\citet{stuckler2008} & Tuberculosis. IMF programs are associated with increased TB incidence and mortality. & Moderate \\
\addlinespace
\citet{stuckler2009} & Global health. IMF programs are associated with weakened health systems. & Review \\
\addlinespace
\citet{stuckler2010} & IMF claims on health. Evidence contradicts IMF claims about its positive impacts. & Review \\
\addlinespace
\citet{stubbs2017} & Social spending. IMF social safeguards are insufficient and programs are associated with lower health spending. & Low \\
\addlinespace
\citet{stubbs2021} & Poverty and inequality. IMF austerity increases inequality and poverty. & Low \\
\addlinespace
\citet{tabassum2023} & Inequality in Pakistan. IMF programs exacerbate income inequality. & Moderate \\
\addlinespace
\citet{tamale2021} & Austerity in COVID-19 loans. Most loans required austerity measures that increase inequality. & Qualitative \\
\addlinespace
\citet{thomson2017} & Maternal and child health. Adjustment programs have a detrimental impact. & Review \\
\addlinespace
\citet{turan2023} & Economic growth. Non-concessional programs have a negative effect on growth. & Moderate \\
\addlinespace
\citet{vreeland2003} & Inequality. IMF programs increase income inequality. & Moderate \\
\addlinespace
\citet{ye1999} & Poverty in Ghana. Structural adjustment reduced poverty through the informal and non-agricultural sector. & Moderate \\
\end{longtable}

\end{document}